\documentstyle[12pt,aasms4]{article}
\begin{document}

\newcommand{\kms}{\mbox{${\rm\,km\,s}^{-1}$}}
\newcommand{\kmsmpc}{\mbox{${\rm\,km\,s}^{-1}{\rm\,Mpc}^{-1}$}}
\newcommand{\pc}{\mbox{$\rm\,pc$}}
\newcommand{\kpc}{\mbox{$\rm\,kpc$}}
\newcommand{\mpc}{\mbox{$\rm\,Mpc$}}
\newcommand{\msun}{\mbox{$M_\odot$}}
\newcommand{\microm}{\mbox{$\rm\,\mu m$}}
\newcommand{\gyr}{\mbox{$\rm\, Gyr$}}
\newcommand{\ang}{\mbox{$\rm\,\AA$}}
\newcommand{\beq}{\begin{equation}}
\newcommand{\eeq}{\end{equation}}

\lefthead{Bak \& Statler}
\righthead{Elliptical Galaxy Shape Distribution}

\title{The Intrinsic Shape Distribution of a Sample of Elliptical Galaxies}
 
\author{Jakob Bak\altaffilmark{1} and Thomas S. Statler\altaffilmark{2}}
\affil{Department of Physics and Astronomy, Ohio University,
    Athens, OH 45701}
\altaffiltext{1}{bak@helios.phy.ohiou.edu}
\altaffiltext{2}{tss@helios.phy.ohiou.edu}

\begin{abstract}

We apply the dynamical modeling approach of Statler
(\markcite{Statler1994b}1994b) to 13 elliptical galaxies
from the Davies and Birkinshaw (\markcite{DB1988}1988) sample of radio
galaxies to derive constraints on their intrinsic shapes and orientations.
We develop an iterative Bayesian algorithm to combine these results
to estimate the parent shape distribution from which the sample was drawn,
under the assumption that this parent distribution has no preferred
orientation. In the process we obtain improved estimates for the
shapes of individual objects. The parent shape distribution shows a tendency
toward bimodality, with peaks at the oblate and prolate limits. Under
minimal assumptions about the galaxies' internal dynamics, 35\% of
the objects would be strongly triaxial ($0.2 < T < 0.8$). However, the parent
distribution is sensitive to the assumed orbit populations in the
galaxies. Dynamical configurations in which all galaxies rotate purely about
either their long or short axes can be ruled out because they would
require the sample to have a strong orientation bias. Configurations in
which the mean motion about the short or long axis is either
``disklike''---dropping off away
from the symmetry planes---or ``spheroidlike''---remaining roughly
constant at a given radius---are equally viable. Spheroidlike rotation in
the long-axis or short-axis tube orbits significantly lowers the abundance
of prolate or oblate galaxies, respectively. If rotation in ellipticals is
generally disklike, then triaxiality is rare; if spheroidlike, triaxiality
is common.

\end{abstract}
 

\section{Introduction}

Over the years a number of attempts have been made to derive the intrinsic
shape distribution of elliptical galaxies from observations (Hubble
\markcite{Hubble1926}1926, Sandage et al.\ \markcite{Sandage1970}1970,
Noerdlinger \markcite{Noerdlinger1979}1979, Marchant \& Olson
\markcite{Marchant1979}1979, Richstone \markcite{Richstone1979}1979,
Binggeli \markcite{Binggeli1980}1980, Binney \& de Vaucouleurs
\markcite{Binney1981}1981, Olson \& de Vaucouleurs \markcite{Olson1981}1981;
for a review see Statler \markcite{Statler1996}1996). Generally the
results of these efforts have been ambiguous, and interest
in the problem waned somewhat in the 1980s. But more recent developments
have sparked renewed attempts to crack this classic chestnut.
Among these developments are the recognition that halo shapes may
serve as a diagnostic of galaxy formation physics (Dubinski \& Carlberg
\markcite{Dubinski1991}1991, Weil \& Hernquist \markcite{Weil1996}1996),
and indications that Hamiltonian chaos, dissipation, or both may either force
triaxial equilibrium configurations to evolve slowly toward axisymmetry
or render them altogether impossible (Dubinski \markcite{Dubinski1994}1994,
Merritt \& Fridman \markcite{Merritt1996}1996, Merritt \& Quinlan
\markcite{Merritt1998}1998).

Studies of central surface brightness profiles using {\sl HST\/} suggest
that the fundamental properties of elliptical galaxies may be bimodally
distributed. There appears to be a dichotomy between high-luminosity,
slowly rotating systems with shallow central cusps and boxy isophotes,
and lower luminosity, rotationally supported systems with steeper cusps and
a tendency for diskiness (Lauer et al.\ \markcite{Lauer1995}1995; see also
Kormendy \& Bender \markcite{Kormendy96}1996). Since rapid rotation and
triaxiality are generally regarded as being incompatible, one might
anticipate a bimodal distribution of triaxialities. Tremblay \& Merritt
\markcite{Tremblay1996}(1996) find that low- and high-luminosity
elliptical galaxies have different distributions of apparent ellipticity,
which would imply different distributions of true shapes. Merritt \&
Tremblay's work joins that of Fasano \& Vio \markcite{Fasano1991}(1991),
Ryden (\markcite{Ryden1992}1992, \markcite{Ryden1996}1996), and Fasano
\markcite{Fasano1996}(1996) as successors to the classical photometric
approaches pioneered by Hubble, Sandage, and others.

However, photometric methods, while effective in constraining the
distribution of overall flattenings, reveal little about the frequency
of axisymmetry {\em vs.\/} triaxiality in the population. Uncovering
this information requires the use of kinematic data and dynamical models
to connect the kinematics to the shape of the gravitational potential.
In the rare cases where well-defined, equilibrium gas disks are present,
emission-line kinematics can yield excellent constraints on the shape
of the potential if one assumes that the gas is on closed orbits (Bertola
et al.\ \markcite{Bertola1991}1991). But for the majority of ellipticals,
methods relying primarily on stellar kinematics are essential. Approaches
of this type were originated by Binney \markcite{Binney1985}(1985) and
enlarged upon by Franx et al.\ \markcite{Franx1991}(1991) and Tenjes
et al.\ \markcite{Tenjes1993}(1993). Statler (\markcite{Statler1994a}1994a,
\markcite{Statler1994b}1994b) introduced major refinements, including
improved dynamical models and a Bayesian approach to model fitting.
This method has great potential to place quite narrow constraints on
the triaxialities of individual galaxies for which very high quality
stellar kinematic data are available. Unfortunately, the number of such
galaxies is still very small, and is likely to increase at only a
modest pace in the short term.

Our goal in this paper is to see what can be learned from the larger
sample of galaxies with stellar kinematic data of less-than-ideal quality
already in the literature. We focus on the Davies \& Birkinshaw
\markcite{DB1988}(1988, hereafter DB) sample of radio ellipticals, all of
which have kinematic data on multiple position angles and are
photometrically well studied. In the process, we extend the statistical
methods of Statler \markcite{Statler1994b}(1994b) and show how to estimate the
parent shape distribution from a sample of galaxies for which the data may be
very inhomogeneous. Our method will thus continue to be generally
applicable as new data are obtained.

In the next section of the paper
we describe the general statistical approach for determining the parent
distribution of a set of intrinsic quantities from measurements of
related, but different, observable quantities, and show the particular
application of this approach to the shape problem. In \S\ 3, we discuss
our treatment of the data and define a subsample of the DB galaxies which
we are able to model reliably. Section 4 presents the results for
the parent distribution, and examines systematic effects relating to unknown
aspects of the stellar dynamics. Section 5 compares our results to those
of previous studies, and \S\ 6 sums up.

\section{Estimating the Parent Distribution}

\subsection{Basic Idea}

In previous papers (Statler \markcite{Statler1994b}1994b,
\markcite{Statler1994c}1994c, Statler et al.\ \markcite{Statler 1999}1999)
we describe a Bayesian approach to inferring the intrinsic shapes
of individual elliptical galaxies, based on
dynamical models that predict the mean radial velocity field (VF)
by solving the equation of continuity for the
stellar ``fluid.'' For triaxial systems with negligible
figure rotation, the streamlines of the mean motion in the main families of
circulating orbits are dictated by the triaxiality of the mass
distribution; thus for a given shape, orientation, and set of boundary
condition parameters describing the internal orbit populations, the
line-of-sight VF can be calculated. The calculation of the VF is very fast,
so the multidimensional parameter space can be adequately explored.

For each set of parameters, we calculate the probability that the observed
VF and surface brightness distribution would, with the known observational
errors, be obtained from the corresponding model. This yields a
multidimensional likelihood function $L(T,c,\Omega,{\bf d})$, where $T$ is
the triaxiality of the total mass distribution, $c$ is the short-to-long
axis ratio of the luminosity distribution, $\Omega = (\theta,\phi)$ is the
orientation of the galaxy, and the vector ${\bf d}$ represents the remaining
dynamical parameters (see \S\ \ref{s.implementation}).
The likelihood is integrated over an assumed prior
distribution in the dynamical parameters ${\bf d}$ and an isotropic prior
distribution in $\Omega$ to give a two-dimensional likelihood
$L(T,c)$.\footnote{See Statler et al.\ \markcite{Statler 1999}(1999) for
a more rigorous formulation of this statement.} To obtain the Bayesian
estimate of the galaxy's shape, $L(T,c)$ is multiplied by a model for the
parent shape distribution and normalized. In most of our previous work,
this model has been a flat distribution, $F(T,c)={\rm const}$, meaning that
we have estimated the shape of each galaxy in isolation.
The likelihood $L_i(T,c)$ for each galaxy $i$
is basically a data point with an error ellipse, i.e., a measurement of $T$
and $c$. Of course, in this case the
errors are strongly non-Gaussian since we work with the actual
probability distribution. The goal now is to combine these measurements for
a sample of galaxies into an estimate of the parent distribution from which
the sample was drawn.

Our algorithm is conceptually simple. We start with a flat
model for the parent distribution, $F(T,c)={\rm const}$. The Bayesian
posterior probability density $P_i(T,c)$ for each galaxy is the
normalized product of $L_i$ with the model parent $F$. (At this stage,
$P_i=L_i$.) We stack the $P_i$'s on top of each other, add them up,
smooth the sum with a nonparametric smoothing spline, and normalize the
result. This gives us an improved model for the parent distribution $F$,
which we multiply by the $L_i$'s and feed iteratively into the same
procedure. Note that, after the first iteration, the statistical estimate
of the shape of each galaxy in the context of the whole sample, $P_i$,
is different from the estimate of its shape in isolation, $L_i$. This
difference arises from the requirement that the sample be drawn from
an isotropic distribution of orientations. Note also that since all
operations are performed on distributions that are already integrated
over $\Omega$, the isotropy of the parent distribution is guaranteed.

A one dimensional toy problem can demonstrate that this algorithm works,
even when the $L_i$ distributions are strongly non-Gaussian. Consider
a set of objects, each with a value of some intrinsic property $X$ between
0 and 1. $X$ is not measurable, but a related quantity $x$ is.
Suppose that for any object an observer has a uniform probability of
measuring any value of $x$ between 0 and $X$. It is easy to show that
a single measurement $x_i$ implies a likelihood $L_i(X)$ that is zero
for $X<x_i$ and proportional to $X^{-1}$ for $X>x_i$. Figure
\ref{f.oned} shows the algorithm in action. The likelihoods $L_i(X)$ for
9 measured $x$ values are shown in Fig.\ \ref{f.oned}$a$. These functions
are multiplied by the initially flat parent distribution (b), summed
(c), and smoothed to produce the new parent (d). The result after
20 iterations (e, solid line) is a decent representation of the true
parent distribution (dotted line). The functions $P_i(X)$ giving the
estimates of the $X$ values for the individual objects (f) differ from the
original $L_i$'s but are consistent with the parent distribution.

\subsection{Statistical Rationale}

The stack-smooth-iterate algorithm is a general
technique that is closely related to Lucy's Method
(Lucy \markcite{Lucy1974}1974) and penalized likelihood (Wahba \&
Wendelberger \markcite{WaW80}1980, Silverman \markcite{Silver1986}1986,
Green \& Silverman \markcite{Green1994}1994).\footnote{Our method is
similar, but not identical, to a well-established approach developed
by Wahba \& Wendelberger \markcite{WaW80}(1980). We are grateful to the
referee for directing us to this important paper.}\ Readers uninterested in
the statistical details are welcome to skip directly to \S\ 2.5.

Imagine a population of objects, each of which possesses
some value of an intrinsic property (or set of properties) $X$, distributed
according to the parent distribution $F_p(X)$. Let there be an observable
quantity (or set of quantities) $x$ which is related to $X$ by a conditional
probability distribution $P(x|X)$.
The distributions are normalized so that $\int dX\, F_p(X) =
\int dx\, P(x|X) = 1$.
>From models, we calculate the likelihood $L(X|x)$ that a given measurement
of $x$ was obtained from an intrinsic value $X$. We assume that the likelihoods
are explicitly normalized so that $\int dX\, L(X|x) = 1$. If the models take
into account all physical effects and measurement errors exactly, then $L(X|x)$
and $P(x|X)$ are mathematically the same function. For
measurements $x$, the estimate of $X$ is
given by the posterior density,
\beq\label{e.posterior}
P(X)={F(X) L(X|x) \over \int dX\, F(X) L(X|x)},
\eeq
where $F(X)$ is the current estimate of the parent distribution.
This is the standard Bayesian approach for individual objects.

The goal is to find the parent distribution $F(X)$ that maximizes the joint
probability of obtaining the measurements $x_i$ for the set of $n$ objects
$i=1,\ldots,n$. The logarithm of this probability is given by
\beq\label{e.likeli}
\ln {\cal P} = \ln \prod_i \int dX\, F(X) P(x_i|X) = 
\sum_i \ln \int dx\, F(X) L(X|x_i).
\eeq
If we write the measurements in terms of a distribution of observables,
$W(x) = \sum_i \delta(x-x_i)$, then we can interpret the integral
\beq\label{e.wmodel}
W_m(x)\equiv \int dX\, F(X) L(X|x)
\eeq
as giving the distribution of observables that would be predicted if the model
parent distribution $F(X)$ were correct. With these definitions,
equation (\ref{e.likeli}) becomes
\beq\label{e.likelihood}
\ln {\cal P} = \int dx\, W(x) \ln W_m(x).
\eeq
If we subtract the constant $\int dx\, W(x) \ln W(x)$ from the quantity in
equation (\ref{e.likelihood}), we get the Lucy $H$-function,
\beq
H_L = \int dx\, W(x) \ln {W_m(x) \over W(x)} .
\eeq
In the statistics literature, $-H_L$ is known as the ``Kullback-Leibler
information distance'' between model and data (Silverman
\markcite{Silver1986}1986).

Lucy's method works to increase $H_L$ (decrease the information distance)
by iteratively applying the rule
\beq
F_{\rm new}(X) = F(X) \int dx\, {W(x) \over W_m(x)} L(X|x).
\eeq
Using equations (\ref{e.posterior}) and (\ref{e.wmodel}), this can be
written as
\beq\label{e.newparent}
F_{\rm new}(X) = \int dx\, W(x) P(X) = \sum_i P_i(X).
\eeq
Thus one iteration of Lucy's method is identical to our scheme of
``stacking'' the posterior densities. In the absence of smoothing, our
approach will seek a parent distribution that maximizes the likelihood
of the observed sample.

For a finite sample, however, the maximum-likelihood parent distribution
will be a set of spikes at the maximum of each $L(X_i|x_i)$, so it is
ill-advised to iterate without a penalty function that enforces smoothness. 
Moreover, for a realistically small sample, there is not a unique $F(X)$
that maximizes $\ln \cal{P}$ unless such a penalty function is present
to lift the degeneracy.
To implement this penalty at each iteration, we regard $F_{\rm new}(X)$,
computed according to equation (\ref{e.newparent}), as a noisy realization
of an underlying smooth function $F^s_{\rm new}(X)$. This function is estimated
using a smoothing spline, and $F_{\rm new}(X)$ is replaced by its smoothed
counterpart.

\subsection{Smoothing Splines and Cross-Validation}

Smoothing splines may be defined in any number of dimensions;
here we summarize the two-dimensional case. This 
discussion is adapted from Green \& Silverman 
\markcite{Green1994}(1994) and Silverman \markcite{Silver1986}(1986).

The estimate of the parent distribution, $F_{\rm new}(X)$, 
is defined on a discrete grid of $n$ values of $X$. (We drop the
subscript ``new'' in what follows for brevity.)
In our case, $X=(T,c)$, and we have $F(T_1,c_1)$,..., $F(T_n,c_n)$, 
from which we want to determine the underlying
smooth function $F^s(T,c)$. For a trial function $g(T,c)$,
a penalized sum of squares of the residuals is given by
\begin{equation}
S(g)=\sum_{i=1}^{n} \{F(T_i,c_i)-g(T_i,c_i)\}^2+
\alpha J(g),
\end{equation}
where $\alpha>0$ is a ``rate of exchange'' between the usual goodness of fit
measure and the penalizing function $J(g)$, given by
\begin{equation}
J(g)=\int \int dT\, dc
\left\{
\left( \frac{\partial^2g}{\partial T^2} \right)^2+
2 \left( \frac{\partial^2g}{\partial T \partial c}\right)^2+
\left( \frac{\partial^2g}{\partial c^2}\right)^2
\right\}.
\end{equation}
The penalizing function measures the rapid variation and departure from
local linearity in $g$.
The functions which minimize $S(g)$ are known as {\em thin plate
splines\/}, which are analogous to natural cubic splines in one dimension.
Algorithms for calculating thin plate splines are implemented in the routines
DTPSS and DPRED in the GCVPACK package
(Bates et al.\ \markcite{Bates1987}1987), which
is available from Netlib.\footnote{http://netlib.org/gcv}

The problem of finding $F^s(X)$ is now reduced to determining a
single parameter $\alpha$. Likelihood cross validation provides a method
for doing this automatically.
The premise is that the best estimate of $\alpha$
should produce the distribution which best predicts all future data points.
Since one is generally not gifted with prescience, one proceeds by removing
one measurement from the sample and calculating the likelihood that that
measurement would be obtained in the parent distribution found from
the other measurements. Repeating this procedure for each measurement in
turn and then averaging the likelihoods yields the cross validation (CV)
score.
 
In our case, the measurements are the individual normalized likelihoods
$L(X_i|x_i)$. We remove the $i$th measurement from our data set and create
a new distribution $F_{-i}(X)$ using the methods above.
For a given value of $\alpha$, the
likelihood that a single $L(X_i|x_i)$ is drawn from the smoothed model
parent distribution $F^s_{-i}(X;\alpha)$ is given by
\begin{equation}
{\cal L}_{-i}=
\ln \left(
\int dX\, F^s_{-i}(X;\alpha)\ L(X|x_i)
\right). 
\end{equation}
Averaging the ${\cal L}_{-i}$'s gives the likelihood cross validation score,
\begin{equation}
CV(\alpha)=n^{-1} \sum_{i=1}^n {\cal L}_{-i},
\end{equation}
and maximizing $CV(\alpha)$ provides the best estimate for $\alpha$. To
ensure uniqueness of the final result, we compute the maximum of $CV(\alpha)$
only once, on the first iteration, and fix the smoothing parameter for all
subsequent iterations to its initial value. In some cases there is not a unique
maximum in $CV(\alpha)$; instead, $CV(\alpha)$ is nearly flat up to some
$\alpha_0$, beyond which it turns over. We set $\alpha$ to its turnover
value, and we see no indication that this choice biases the results.

\subsection{Implementation\label{s.implementation}}

Our numerical implementation follows from that described in Statler
\markcite{Statler1994b}(1994b). The treatment of individual galaxies
is essentially the same, except for details noted in \S\ 3
below. The grid of dynamical parameters used in the models is also the
same as in the earlier work; to aid the reader in \S\ 4 we give a brief
overview here.

The models assume that (1) rotation of the figure (i.e., tumbling) is
negligible; (2) short-axis tube and long-axis tube mean motions can be
represented by confocal streamlines (Anderson \& Statler
\markcite{Anderson1998}1998); (3) the luminosity density $\rho_L$ is
stratified on similar ellipsoids, $\rho_L (r,\theta,\phi) = \bar{\rho}_L(r)
\rho_L^\ast(\theta,\phi)$; and (4) the velocity field obeys a ``similar
flow'' ansatz outside the tangent point for a given line of sight,
${\bf v}(r,\theta,\phi) = \bar{v}(r) {\bf v}^\ast(\theta,\phi)$.
The last two assumptions are needed for projecting the models.
The results are insensitive to the accuracy of these assumptions
as long as $\bar{\rho}_L(r)$ and $\bar{\rho}_L(r) \bar{v}(r)$ decrease faster
than $r^{-2}$. This requirement limits the validity of the models to regions
where the rotation curve is not steeply rising. As a further simplification
we adopt power laws for the luminosity density and the
velocity scaling law: $\bar{\rho} \sim r^{-k}$ and $\bar{v}(r) \sim r^{-l}$.
The index $k$ is determined from surface photometry,
and we nominally adopt $l=(0,\pm\onehalf)$, omitting the
$l=-\onehalf$ case when $k\leq 2.5$. It turns out that the results
are not very sensitive to either of these parameters.

Remaining properties of the phase space distribution function are described
by a scalar constant $C$ and a function of one variable $v^\ast(t)$.
These parameters describe the mean velocity across the $xz$ plane on one
fiducial shell, which in turn determines the velocity field over the whole
shell once the triaxiality $T$ and the luminosity density are specified.
The ``contrast'' $C$ is defined as the ratio of the $y$ component of the
mean velocity on the $x$ axis to that on the $z$ axis, on the fiducial
shell. The function $v^\ast(t)$ gives the angular dependence of the mean
velocity across the $xz$ plane on the fiducial shell. The variable $t$ is a
rescaled polar angle, given, for spherical shells, by
\beq
t = \left\{ \begin{array}{ll}
        2 - {\sin^2 \theta \over T}, & \theta < \sin^{-1}\sqrt{T}, \\
        {\cos^2 \theta \over 1-T}, & \theta > \sin^{-1}\sqrt{T},
\end{array} \right.
\eeq
where $\theta$ is the usual polar angle. The relation for ellipsoidal shells
is given in \S\ Section 3.1 of Statler \markcite{Statler1994b}(1994b). By
definition, $v^\ast(0)=C$ and $v^\ast(2)=1$.

The model grid comprises 8 different assumptions for the variation of $C$ with
intrinsic shape. In four of these $C$ is constant: $C=0$ (long-axis
tube dominated), $0.5$, $1$, and infinite (short-axis tube dominated).
Four more functional forms for $C(T,c)$ are introduced to mimic
certain self-consistent models, and are given in equations (11) -- (14)
of Statler \markcite{Statler1994b}(1994b). The function $v^\ast(t)$ is
taken to be either piecewise-constant or piecewise-linear in each of the
intervals $[0,1)$ and $(1,2]$ (in the linear cases dropping to zero at
$t=1$). This function describes how the mean rotation speed in each of
the tube orbit families declines away from the symmetry plane that
contains its parent orbits. For example, the mean rotation in the
Galaxy drops with height above the disk plane as one moves into the more
pressure-supported halo. At the other extreme, a maximally rotating
isothermal sphere has constant rotation speed at all latitudes.
Accordingly, we refer to linear $v^\ast(t)$ as ``disklike'' rotation, and
constant $v^\ast(t)$ as ``spheroidlike'' rotation. A model can be disklike
or spheroidlike in either short-axis or long-axis tubes. One should avoid
the impression that disklike rotation necessarily implies a
two-component structure; in an oblate disklike model, the mean rotation speed
$45\arcdeg$ up from the equatorial plane is half of the in-plane value,
a much gentler transition than in a genuine disk-halo system.

The likelihoods $L_i(T,c)$ for each galaxy are computed
on a $20 \times 20$ rectangular grid on the intervals $0 \leq T \leq 1$
and $0.4 \leq c \leq 1$. Smoothing a function over a finite domain
creates problems near the edges unless suitable boundary conditions are
imposed to minimize this effect. The thin plate spline does not impose 
any strict boundary conditions but rather sees the area outside of the 
boundaries as lacking information. The penalizing 
function $J(L_i(T,c))$ 
is therefore the only part to contribute to the penalized sum of
squares of the residuals, forcing 
the function $L_i(T,c)$ to be flat outside the
boundaries\ (see section 2.3). In practice this has the effect 
of biasing $L_i(T,c)$ towards closed contours and reduced variability near
the edges (Green \& Silverman \markcite{Green1994}1994). The effect however is 
limited and our results show little if no evidence of it.

We find that, in practice, convergence of the parent distribution can be
rather slow, as peaks grow at the expense of valleys that sink toward
zero. With our sample of only 13 objects, iterating until a stringent
convergence criterion is satisfied may be dangerous. In order to be
conservative in our conclusions regarding the frequency of triaxiality,
we stop iterating when the maximum
fractional change in $F(T,c)$ per iteration falls below 10\%. Typically
this occurs after about 7 iterations.

\section{Data}

\subsection{Kinematics}

All of the galaxies modeled are taken from the sample of radio ellipticals
for which DB obtained multiple position angle rotation curve measurements.
The sample contains more E3--E4 and fewer E0 galaxies than the general
population and, as DB point out, includes an overabundance of ``unusual''
objects. Where appropriate we have supplemented the DB data with data from
Franx et al.\ (\markcite{Franx1989}1989),
Binney et al.\ (\markcite{Binney1990}1990),
Bender et al.\ (\markcite{Bender1994}1994) and 
Fried \& Illingworth (\markcite{Fried1994}1994).

The published rotation curves are first oriented to match our convention that
radii west of north are positive. Since the models assume that the rotation
curves are antisymmetric, we fold the profiles about the center of the galaxy
to reduce the formal errors in the average rotation velocity. For each
individual galaxy we approximate by eye the radius at which the rotation
curve flattens and use the data outside of this in the models. At large
radii the kinematic data become unreliable
for reasons which vary from galaxy to galaxy (see section 3.3). We
therefore set an outer radius beyond which we discard the data. We average
the data points which are left between the inner and outer radii on each
PA, weighted by the inverse square of the published errors. The
uncertainty associated with the average is taken to be the
$(1/{\sigma^2})$-weighted standard deviation following
Statler (\markcite{Statler1994c}1994c).
The inner and outer radii and the adopted mean velocities are given
in columns 4, 5 and 9, respectively, in Table \ref{tbl-1}.

\subsection{Photometry}

With the exception of the data for NGC 4839, all of the photometry is drawn
from Peletier et
al.\ (\markcite{Peletier1990}1990), who tabulate the ellipticity, major
axis PA and surface brightness as functions of radius. Similiar photometry for 
NGC 4839 is drawn from Joergensen et al.\ (\markcite{Joergensen1992}1992).
For each galaxy
the adopted major axis position angle is the average between the inner
and outer limiting radii. The ellipticities are determined by taking
the unweighted mean in the same interval, with the standard deviation serving
as the uncertainty. The adopted mean ellipticities and major axis PAs are
given in columns 2 and 3 of Table \ref{tbl-1}.

The slope of the surface brightness profile is calculated by differentiating
numerically. The surface brightness slope is then deprojected into a volume
brightness slope ($k$) by adding 1. Although this is strictly valid only
for pure power-law profiles, it is fine for our level of approximation.
For most galaxies the logarithmic slope of the surface brightness
profile is not constant in the relevant intervals, and so two values are used
that span the ranges of $k$.  We compute all of the models using both
values. The spanning values of $k$ for all of the galaxies are 
in columns 6 and 7 of Table ~\ref{tbl-1}.

\subsection{Notes on Individual Galaxies}

Of the 14 galaxies in the DB sample,
four, NGC 1600, NGC 4374, NGC 4636 and NGC 4839, do not
show any significant rotation at DB's level of accuracy.
We therefore model them using only their photometric data. A fifth object,
NGC 4278, does show significant 
rotation but is not used. It shows a $20\arcdeg$
isophotal twist between $20\arcsec$ and $60\arcsec$ and a drop in
rotation velocity to zero outside of $20\arcsec$ that conflicts with our
assumption that the rotation curve is flat at large radii. It could be
modeled but a more sophisticated method involving fitting at multiple radii
would be required.

Details of how we have handled the data for the remaining sample of 13 galaxies
are as follows:
 
{\bf NGC 1600, NGC 4374, NGC 4636 and NGC 4839.}
Because of the lack of any significant rotation in these galaxies,
photometric data alone is used to estimate their shape likelihood
distributions. The triaxialities of these four galaxies are
therefore poorly constrained. They are included on the grounds that
omitting them could bias our results away from strongly triaxial
systems, most of which are probably slowly rotating.
In each case, the data are averaged from the center out to the
largest radius for which kinematic data is available (see Table \ref{tbl-1}).
The ellipticity of NGC 1600, NGC 4374 and NGC 4636 varies by about
$0.10$ in this interval, but the ellipticity of NGC 4839 rises steadily from
$0.20$ near the center to $0.50$ at $32\arcsec$. 

{\bf NGC 315.}
The turnover radius of the rotation curve is easily located at $5\arcsec$ by
inspection. At $27\arcsec$ on PA 40 the rotation velocity is more than
$3\sigma$ from the mean. Since we do not know if this is a real effect,
we eliminate the total of six data points outside of $20\arcsec$.
 
{\bf NGC 741.}
The relatively low surface brightness of this galaxy results in large
uncertainties in the kinematic data. Nonetheless, its six position
angle measurements of the rotation curve make it very attractive to model.
Outside of $15\arcsec$ some rotation appears on PA 10 and PA 40.
Although it is not at all clear that the turnover radius of the rotation
curve has been found, all of the data from $15\arcsec$ to the outermost
data point at $30\arcsec$ are used.
 
{\bf NGC 1052.}
The data for this galaxy is the best for any in the sample. DB
present kinematic data on four PAs, Binney et
al.\ (\markcite{Binney1990}1990) on
the major and minor axis and Fried \&
Illingworth (\markcite{Bender1994}1994) on the major axis.
The turnover radius
of the rotation curve is easily seen to be at $15\arcsec$ for this galaxy.
A velocity difference of almost 50 km/sec between the southern and northern
parts of the galaxy outside of $37\arcsec$ on PA 117 and PA 164 in both the
DB and Binney et al.\ (\markcite{Binney1990}1990) data sets imply
that NGC 1052 may not be antisymmetric outside of this radius as is
assumed in our models. The total of 11 questionable points outside
of $35\arcsec$ are therefore eliminated.
Fried \& Illingworth\ (\markcite{Bender1994}1994)
measure the rotation curve to be flat out to $40\arcsec$ on PA 117.

{\bf NGC 3379.}
The turnover radius of the rotation curve appears at about $15\arcsec$ so
only data from this radius to the outermost data point at $34\arcsec$ is used.
 
{\bf NGC 3665.}
The rotation curve is flat from inside of $5\arcsec$ to close to
$30\arcsec$, but there is a discontinuity
at $10\arcsec$ where the ellipticity suddenly drops from $0.35$
to almost zero. The ellipticity then slowly rises to approximately $0.2$
at $15\arcsec$ outside of which it is constant. There is also a $10\arcdeg$
isophotal twist in the same range. Only data outside of $15\arcsec$ is used.
 
{\bf NGC 4261.}
DB provide kinematic data on four position angles 
out to $55\arcsec$. The major and minor axis data are supplemented with data
from Bender et al.\ (\markcite{Bender1994}1994),
who place their slits $4\arcdeg$
from those of DB.  The average of the slit
positions of the two papers is therefore used when combining the two datasets.
Although this does introduce some error into the data it is very minor
compared to the uncertainty in the velocity measurements.
This galaxy is clearly a minor axis rotator with a
turnover radius of the rotation curve at $20\arcsec$.

{\bf NGC 4472.}
The turnover radius of the rotation curve on all the PAs is at
approximately $25\arcsec$. Between $3\arcsec$ and $30\arcsec$ the
ellipticity of NGC 4472 increases from $0.06$ to $0.17$ but is constant
outside this range, so only data points beyond $30\arcsec$ are used. Outside of
$60\arcsec$ two data points on PA 160 are more than $4\sigma$ from the average
so we discard the points outside of this radius on all position angles.

{\bf NGC 4486.}
This galaxy's slow rotation makes the  errors relatively
large in the velocity measurements, but outside of $20\arcsec$ the data is
statistically consistent with a flattening of the rotation curve out to the
last datapoint at $60\arcsec$. The data between these radii are therefore used.

{\bf NGC 7626.}
This is another slow rotator. The only significant rotation is
on the minor axis. Outside of $20\arcsec$ the rotation curve seems to
reverse, but the errors are so large that the reversal is not
statistically significant. We model the data only inside $20\arcsec$.
A turnover radius in the rotation curve on the major axis at approximately
$5\arcsec$ sets the inner radius.

\section{Results}

\subsection{The ``Maximal Ignorance'' Shape Distribution\label{s.ignorant}}

As in previous papers, we take the result from an unweighted combination of
all models to represent the case of ``maximal ignorance,'' i.e., minimal
assumptions as to the character of the internal dynamics. The parent
shape distribution after seven iterations is shown in Figure
\ref{f.ignorant}$a$. The distribution is plotted in terms of $T$ and $c$ such
that oblate spheroids, prolate spheroids, and spheres lie, respectively,
along the right, left, and top margins. This distribution is
bimodal, dominated by one group of nearly oblate, moderately flattened
systems and a second group of rounder, nearly prolate systems. The valley
between the two peaks represents a dearth of very triaxial galaxies.
The bimodality is almost entirely a consequence of the kinematic data;
to illustrate, we show in Figure \ref{f.ignorant}$b$ the result
obtained from photometry alone, ignoring the kinematics. As discussed in
the Introduction, photometry is effective in constraining the overall
flattening distribution but reveals little about triaxiality.

A more succinct description of the frequency of triaxiality in this
distribution comes from the one-dimensional distribution $F(T)$, obtained by
integrating Figure \ref{f.ignorant}$a$ over $c$. The result is shown in
Figure \ref{f.ignorantoned}$a$. We somewhat arbitrarily set boundaries at
$T=0.2$ and $T=0.8$ to delineate ``nearly oblate,'' ``triaxial,'' and
``nearly prolate'' regions. By this definition, the maximal ignorance
distribution is 47\% nearly oblate, 18\% nearly prolate, and 35\% triaxial.
 If we continue iterating beyond
our nominal stopping criterion, the triaxial fraction decreases further,
so we can take this as a conservative estimate of the rarity of triaxial
systems implied by our subsample of the DB galaxies. The result is
influenced somewhat by the four galaxies without kinematic data; if these
objects are omitted, the fractions change to 55\% oblate, 25\% prolate,
and 21\% triaxial.

We can obtain some measure of whether the DB subsample is representative
of the elliptical galaxy population at large by calculating the expected
ellipticity distribution for a randomly-oriented population drawn from
the inferred parent. We plot this as the smooth curve in Figure
\ref{f.ignorantoned}$b$, compared with the observed ellipticity
distribution from Ryden (\markcite{Ryden1992}1992).
The two distributions are similar, though our predicted distribution
contains a slight excess of very round galaxies. A Kolmogorov-Smirnov (KS) test
implies a 14\% probability that the observed sample 
was drawn from our distribution. However, the KS probability can be affected
by details of how the mean ellipticities are defined. The
ellipticities tabulated by Ryden are weighted by luminosity, whereas we
exclude data from the brightest parts of the galaxies. Applying a
systematic shift as small as $\Delta \epsilon = 0.019$ to our expected
distribution would increase the KS probability to 99\%.
We conclude that our maximal ignorance
parent distribution is consistent with the ellipticities
of the general population of elliptical galaxies.

The final posterior densities describing the shapes of the individual galaxies
in the sample with rotation data are shown 
in Figure \ref{f.montage}. Some well-known objects
are found to have well constrained triaxialities;
NGC 1052, NGC 3379, and NGC 4472 are probably oblate or nearly so.
The famous minor-axis rotator NGC 4261, not surprisingly, turns out to
be most likely prolate, though there are oblate models not excluded
at the $2\sigma$ level. The shapes of other objects are not as well
constrained, and bimodal posterior densities are less a consequence of
the kinematic data for the individual galaxies than a reflection of the
parent distribution.

\subsection{Dynamical Configurations That Can Be Ruled Out}

Just as the marginal posterior densities describing the shape of each
galaxy (Fig. \ref{f.montage}) can be computed for a given parent
distribution, we can compute marginal densities describing the orientation
of each galaxy according to
\beq\label{e.piofomega}
P_i(\Omega)=\int dT \int dc\, {1 \over 4\pi} F(T,c) L_i(T,c,\Omega).
\eeq
The 4-dimensional likelihoods $L_i(T,c,\Omega)$ are obtained by integrating
the original likelihood function $L_i(T,c,\Omega,\bf{d})$ over the dynamical
parameters. The factor $1/4\pi$ reflects the assumed isotropy of the
parent distribution; in other words we have assumed that the 4-dimensional
parent has the form $F(T,c,\Omega) = F(T,c)/4\pi$. We could, in fact,
have worked our whole procedure in 4 dimensions instead of 2. Had we done
so, isotropy of the parent would have been imposed at each iteration
by explicitly smoothing away all of the $\Omega$ dependence from the stacked
$P_i(T,c,\Omega)$ functions. One would expect, for a plausible set of
models leading to a plausible parent distribution, that the stacked
$P_i$'s should have an $\Omega$ dependence not too far from isotropic,
before it is smoothed away. This gives us an important consistency check:
the sum, $\Sigma_i P_i(\Omega)$, of the final posterior densities from
equation (\ref{e.piofomega}) ought to be reasonably flat. Even though
the parent distribution is, by construction, isotropic, there is no
guarantee that the {\em sample\/} is isotropic. If we find a strong
orientation bias in the sample despite assuming an isotropic parent, this
constitutes a contradiction and signals a false assumption.

Two applications of this test are shown in the bottom four panels of
Figure \ref{f.ignorant}.  Figure \ref{f.ignorant}$c$ shows the parent
distribution derived under the assumption that all galaxies rotate about
their intrinsic long axes ($C=0$). Most objects are close to oblate and
quite flat, with a small but significant fraction of rounder, triaxial
systems. This is clearly different from the maximal-ignorance
distribution in Figure \ref{f.ignorant}$a$. However, this case can be
ruled out by the orientation distribution of the sample, shown in Figure
\ref{f.ignorant}$e$. For the galaxies all to be long-axis
rotators, we must be seeing them in nearly the same
orientation; the line of sight lies inside one of two 
$45\deg$-wide cones for about 40\% of the sample.

A better quantitative measure of the orientation bias is the rms
deviation of $\Sigma_i P_i(\Omega)$ from perfect isotropy, normalized
to unit mean; we refer to this as the {\em sample anisotropy\/}, $A_s$.
For the case in Figure \ref{f.ignorant}$e$, $A_s=1.17$. Table \ref{tbl-2}
gives $A_s$ values for the distributions calculated using various
subsets of the dynamical models. Unfortunately, it is not straightforward
to link a value of $A_s$ with a confidence limit. The
expected $A_s$ distribution for an ensemble of random isotropic samples
depends on the forms of the individual $P_i$'s, which depends on both
data and models. We can make a very rough correspondence to an easier
statistical problem if we imagine that each $P_i(\Omega)$ simply marks
a fraction $f$ of the sphere as allowed and a fraction $1-f$ as
excluded. The $P_i$'s are then $n$ patches thrown down at random onto the
sphere. At a random point on the sphere, the number $m$ of overlapping patches
is given by a binomial distribution. In Table
\ref{tbl-2}, we find that, except for the $C=0$ and $C=\infty$ cases,
all of the models using the kinematic data hover around $A_s \approx
0.2$. This would follow from the binomial distribution for $f=0.63$,
which is a not-unreasonable characterization of the $P_i$'s. We calculate
that, if $0.2$ is the expected $A_s$ for a random sample of 13 objects and
if $A_s$ is distributed as in the patch problem, we can reject
cases with $A_s > 0.40$ at $99\%$ confidence and cases with $A_s > 0.48$
at $99.9\%$ confidence. A more realistic simulation with 9 patches that
exclude 50\% of the sphere and 4 that exclude only 10\% gives very
similar results. Thus the hypothesis that elliptical galaxies rotate
about their long axis is firmly ruled out. Of course, this is neither
a particularly surprising nor new result; Binney \markcite{Binney1985}(1985)
reached the same conclusion from essentially the same data.

Figure \ref{f.ignorant}$d$ shows the parent distribution under the
assumption that all objects rotate around their intrinsic short axes
($C=\infty$, also known as ``zero intrinsic misalignment''). Here, most
objects are triaxial, again very different from the maximal ignorance
result. Figure \ref{f.ignorant}$f$ shows the orientation distribution
for the sample, which has $A_s=0.43$. The assumption of zero intrinsic
misalignment for all systems is excluded at approximately the
$99.6\%$ confidence level. This result
differs from that of Franx et al.\ \markcite{FIZ1991}(1991), who were
able to reproduce the observed distribution of ellipticities and
kinematic misalignment angles\footnote{For galaxies with only
major and minor axis kinematics, the misalignment angle is
$\tan^{-1}(v_{\rm minor}/v_{\rm major})$.}
with a family of triaxial models rotating
about their short axes. We have not explored the source of this
disagreement in depth. While it may be due simply to our smaller
sample, we suspect that the models with which Franx et al.\ can fit
galaxies with large kinematic misalignments fail on more detailed
comparison with multi-position-angle data.

\subsection{Dynamical Configurations That Cannot Be Ruled Out}

Of the parent distributions we have derived from various subsets of the
dynamical models, we find no other cases that can be ruled out on the basis
of the $A_s$ values. Some of the unexcludable cases nonetheless differ
significantly from the maximal ignorance distribution. Of particular
interest are the cases in the last four rows of Table \ref{tbl-2}, for
which the derived parent distributions are shown in Figure \ref{f.xmodels}.
These distributions differ only in whether the mean rotation
is assumed to be disklike or spheroidlike
(see \S\ \ref{s.implementation}). The parent distribution is more
sensitive to this assumption than to any of the other
dynamical parameters, save for the cases already ruled out above.

Figure \ref{f.xmodelsT} shows the triaxiality distributions for these
four cases, indicating the fraction of nearly oblate, nearly prolate,
and triaxial systems as defined in \S\ \ref{s.ignorant}.
The prevalence of axisymmetric systems over triaxial ones is
significantly affected by the rotation characteristics, in the sense
that axisymmetry becomes less common if rotation is more spheroidlike. 
Moreover, the fractions of nearly prolate and nearly oblate objects are
largely determined, respectively, by the character of the rotation in the
long-axis and short-axis tubes. The peak at the prolate limit seen in the
maximal ignorance result disappears entirely if the rotation in the
long-axis tubes is spheroidlike. The dominant peak at the oblate limit
is lowered by nearly a factor of two if the short-axis tube rotation is
spheroidlike rather than disklike.

We consider this the most important result in this paper:
{\em if rotation in ellipticals is generally disklike, then
triaxiality is rare; if spheroidlike, triaxiality is common.\/} It
follows that understanding the shapes of elliptical galaxies is closely
linked with understanding whether weak disks are common structural
components. It also follows that a physical understanding of what
conditions during formation are likely to impose disklike or spheroidlike
rotation on a hot stellar system would be extremely valuable.

\section{Discussion}

\subsection{Previous Results on the Shape Distribution}

A number of attempts have been made in the past to determine the 
parent intrinsic shape distribution of elliptical galaxies, mostly using
photometry alone. It is interesting to see how our maximal-ignorance result
compares to some of these.

Ryden (\markcite{Ryden1992}1992) fits a parent distribution to a sample of
171 measured ellipticities by letting the distribution assume the form of
a circular Gaussian in axis ratio space. She finds a best-fit center to
the distribution at $b=0.98$, $c=0.69$, implying that the most common
shape is nearly oblate. The distribution is wide, however, with $61\%$
of galaxies having a triaxiality between $0.2$ and $0.8$, compared to
$35\%$ for our sample. This difference may be attributable to Ryden's
assumption of a single peak; our method shows that the distribution may
be bimodal. When recast in terms of $(T,c)$, Ryden's distribution has a
short-to-long axis ratio expectation value $\langle c \rangle=0.68$,
similar to our value of $0.71$.  

Lambas et al.\ (\markcite{Lambas1992}1992) take a similar approach using 
2135 measurements of ellipticities from the APM Bright Galaxy Survey.
Using a Monte Carlo technique they find the elliptical  
Gaussian in axis ratio space which best reproduces their observations.
Their results are remarkably different both from ours and from Ryden's.
They find the center of their distribution at a flattening $c=0.55$ with
a width of $0.2$ in that dimension, implying that $30\%$ of ellipticals 
have $c<0.4$. The main reason for this difference is an excess of flat
galaxies in their sample. Only $2\%$ of the galaxies in Ryden's
(\markcite{Ryden1992}1992) sample have an apparent ellipticity
$\epsilon > 0.6$, but the APM sample has $30\%$ to $40\%$ in that range.
Lambas et al.\ do not offer an explanation for this apparent inconsistency
with previous photometric studies of elliptical galaxies. Conceivably a
large S0 contamination could be the cause.

A nonparametric, maximum-entropy shape distribution for the Ryden
(\markcite{Ryden1992}1992) ellipticity sample is derived using a modified
Lucy's method by Statler (\markcite{Statler1994a}1994a). He finds a
rather broad distribution in triaxiality, with $47\%$ of the galaxies
having $T<0.5$ compared with $70\%$ in our distribution.

Using the same data as Statler (\markcite{Statler1994a}1994a), Tremblay and
Merritt (\markcite{Tremblay1995}1995) use a nonparametric maximum penalized
likelihood estimator to derive the maximum-entropy shape distribution.
They find that it is weakly bimodal and weighted towards oblate figures.
Our distribution is significantly more bimodal and predicts fewer triaxial
galaxies.

Using the same technique,
Tremblay and Merritt (\markcite{Tremblay1996}1996) 
estimate the parent distribution from a sample of 220 ellipticities.
They assume that all galaxies have the same triaxiality and then
proceed to calculate the distribution of intrinsic flattenings $c$. 
They find that a pure oblate or prolate distribution is inconsistent with 
the available data and that a division of intrinsic flattenings exists
between bright and faint galaxies with peaks at $c=0.75$ and $c=0.65$
respectively.  All our galaxies are bright and therefore our 
expectation value of $c=0.71$ agrees well with theirs.

Although the above studies, with the exception of Lambas et
al.\ (\markcite{Lambas1992}1992), give similar results for the axis
ratio $c/a$, none is
able to put any real constraints on triaxiality, even when large 
samples are used. This demonstrates the need to include kinematic 
data in the models. Franx, Illingworth and de Zeeuw (\markcite{Franx1991}1991)
attempt to address this need by including the misalignment between the
photometric and kinematic axes in their models. Studying a sample of 38 
ellipticals, they conclude that a wide variety of distributions are 
consistent with the data, including ones similar to ours with both an
oblate and a prolate peak.

\subsection{Previous Results for Individual Galaxies}

Some of the individual galaxies in our sample have been modeled previously.
Statler (\markcite{Statler1994c}1994c) treats NGC 3379 using essentially
the same data and methods applied here, except that the galaxy is fit in
isolation, using a flat parent distribution. The result is that flattened
nearly oblate shapes or rounder triaxial configurations are allowed by the
data. Compared with this earlier result, the posterior density shown in
Figure \ref{f.montage} is more constrained toward small $T$
due to the preference for near axisymmetry in the parent distribution.

Some objects in our sample have available additional kinematic or
morphological constraints which are not included in our models. The
best-studied example is NGC 1052, which has been modeled by Binney et
al.\ (\markcite{Binney1990}1990), Tenjes et al.\ (\markcite{Tenjes1993}1993), 
and Plana and Boulesteix (\markcite{Plana1996}1996). This galaxy has the
best constrained shape in our sample; Figure \ref{f.montage} shows only
a small permitted region around the oblate spheroid with $c=0.63$.
The small triaxiality supports the use of axisymmetric models by
Binney et al. \markcite{Binney1990}(1990) to constrain the phase space
distribution function. Applying the Jeans equation to the observed
surface photometry and comparing the predicted velocity dispersion and 
azimuthal streaming to the observed kinematics, they 
find that NGC 1052 is consistent with a two integral distribution 
function. Tenjes et al.\ \markcite{Tenjes1993}(1993), using the method
of Franx, Illingworth and de Zeeuw \markcite{Franx1991}(1991)
and the presence of a gas disk to constrain the viewing angles,
find that, depending on the specific kinematic model used, $c$ lies
between $0.4$ and $0.6$ and the triaxiality is well constrained between
$0.56$ and $0.61$. These values imply a very highly triaxial galaxy,
and lie well outside of our $95\%$ highest posterior density region.
Using similar methods Plana \& Boulesteix (\markcite{Plana1996}1996)
calculate the triaxiality to be $0.48$ with a flattening of $0.5$. This is
much flatter and more triaxial than our result. It is possible that including
orientation constraints from the gas would alter our derived shape.
However, the results of Tenjes et al.\ \markcite{Tenjes1993}(1993) and
Plana and Boulesteix (\markcite{Plana1996}1996) are very 
sensitive to the orientation of the disk, and consequently to
assumptions about its intrinsic flatness and circularity; even a 
small error here could change their results dramatically.

In a study similar to that of Binney et al.\ (\markcite{Binney1990}1990), 
Van der Marel et al.\ (\markcite{Marel1990}1990) model the distribution
functions of NGC 3379, NGC 4261, and NGC 4472. Their use of oblate axisymmetric
models for NGC 3379 and NGC 4472 is supported by our results for these
galaxies. For NGC 4261 they fit the observations to a prolate model 
with $c=0.59$, which is consistent with our triaxiality 
estimate and is within our $95\%$ highest posterior density region.
	
At the risk of disappointing the reader, we have avoided discussing the
orientations of individual galaxies in the sample. This is, admittedly,
counter to the original motivation of DB, which was to determine if there is
any relationship between the orientations of the galaxies and their radio jets.
Although we have calculated orientation constraints for each of the galaxies,
a full discussion of this topic would of necessity be lengthy, and is
outside the scope of this paper. We will deal with this issue in a future 
publication.
 
\section{Summary and Conclusions}

By combining photometric and kinematic data with dynamical models using the
method of Statler (\markcite{Statler1994b}1994b), we have derived constraints
on the intrinsic shapes and orientations of 13 ellipticals from the 
Davies and Birkinshaw (\markcite{DB1988}1988) sample of radio galaxies. 
Using an iterative Bayesian approach we have then combined those results
to estimate the parent shape distribution from which they were drawn,
under the assumption that this parent distribution has no preferred
orientation. In the process we have obtained improved constraints on the
shapes of the individual objects.

We have found that the parent shape distribution shows a tendency toward
bimodality, with peaks at the oblate and prolate limits. In the distribution
derived under minimal assumptions about the galaxies' internal dynamics, only
about a one-third of the objects would be strongly triaxial ($0.2 < T <
0.8$). However, the parent distribution does depend on dynamical
assumptions. Some of these assumptions can be ruled out because they
would require the sample to have a strong orientation bias;
configurations in which all galaxies rotate purely about either their
long axes or their short axes can be excluded on these grounds. On the
other hand, configurations in which the mean motions in the short-axis
and long-axis tube orbits are either disklike---dropping off away from the
symmetry planes---or spheroidlike---staying approximately constant at a
given radius---cannot be distinguished at this point. Whether the
rotation is disklike or spheroidlike has a strong effect on the inferred
shape distribution. Spheroidlike rotation in the long-axis or short-axis
tubes, respectively, significantly reduces the fraction of nearly prolate or
nearly oblate galaxies; bimodality is completely eliminated if the long-axis
tubes are spheroidlike and the short-axis tubes disklike. In a nutshell,
{\em if rotation in ellipticals is generally disklike, then
triaxiality is rare; if spheroidlike, triaxiality is common.}

This inferential link between diskiness and axisymmetry complements
the intuitive physical notion that the two ought to go hand in hand.
There is evidence from the width of the Tully-Fisher relation that 
the disks of spiral galaxies are very nearly circular (Franx \& de Zeeuw
\markcite{FdZ92}1992), and indications from numerical experiments that
growing even a weak disk in a triaxial halo can render the latter
axisymmetric (Dubinski \markcite{Dubinski1994}1994). Whether weak
disks in elliptical galaxies are detectable is another long-standing issue
receiving renewed attention (Magorrian \markcite{Mag99}1999). High-accuracy,
multi-position-angle
kinematic mapping may be able to reveal hidden disks, but the expected
signatures are subtle. Some support is lent to the possibility that weak
disks may be common by the kinematic similarities that the ``standard
elliptical'' NGC 3379 shares with the S0 galaxy NGC 3115 (Statler \&
Smecker-Hane \markcite{SSH99}1999). Theoretically, however,
the origin of these particular kinematic features is not understood. As we
have stressed, a physical understanding of the processes that
may establish disklike or spheroidlike rotation in a hot stellar system
is sorely needed.

\acknowledgments

We are indebted to Barbara Ryden and the referee, David Merritt, for
numerous constructive comments.
This work was supported by NASA Astrophysical Theory Program Grant
NAG5-3050 and NSF CAREER grant AST-9703036.

\begin{deluxetable}{rrrrrrrrrl}
\footnotesize
\tablecaption{Observational data used in models \label{tbl-1}}
\tablewidth{0pt}
\tablehead{
\colhead{Galaxy} & \colhead{$\epsilon$} & 
\colhead{PA$_{major}$} &
\colhead{R$_{min}\arcsec$} & \colhead{R$_{max}\arcsec$} &
\colhead{$k_1$} & \colhead{$k_2$} &
\colhead{PA\tablenotemark{a}} & \colhead{$v$} &
\colhead{Reference\tablenotemark{c}}
}

\startdata
NGC 315 &0.27$\pm$0.02  
&43
&5 &20
&9/4 &11/4
&178 &18$\pm$10 &1 \cr
& & & & & & &57 &29$\pm$11 &1 \cr
& & & & & & &87 &30$\pm$12 &1 \cr
& & & & & & &118 &20$\pm$10 &1 \nl

NGC 741 &0.17$\pm$0.01 
&89
&15 &30
&9/4 &11/4
&11 &$-$54$\pm$10 &1 \cr
& & & & & & &41 &$-$34$\pm$ 4 &1 \cr
& & & & & & &71 &$-$13$\pm$32 &1 \cr
& & & & & & &101 &12$\pm$12 &1 \cr
& & & & & & &131 &$-$12$\pm$17 &1 \cr
& & & & & & &161 &$-$9$\pm$23 &1 \nl
 
NGC 1052 &0.31$\pm$0.02 
&115
&15 &40
&10/4 &12/4
&3 &13$\pm$11 &1,3,5\cr
& & & & & & &48 &53$\pm$ 5 &1 \cr
& & & & & & &92 &97$\pm$12 &1,3 \cr
& & & & & & &139 &43$\pm$ 7 &1 \nl

NGC 1600 &32$\pm$0.03 
&13
&0 &27
&\nodata &\nodata
&\nodata &\nodata &1\nl

NGC 3379 &0.11$\pm$0.02 
&70
&15 &50
&9/4 &11/4
&60 &$-$37$\pm$12 &1 \cr
& & & & & & &89 &$-$53$\pm$12 &1,2 \cr
& & & & & & &129 &$-$37$\pm$14 &1 \cr
& & & & & & &179 &3$\pm$18 &1,2 \nl

NGC 3665 &0.23$\pm$0.005 
&25
&15 &30
&10/4 &11/4
&11 &27$\pm$12 &1 \cr
& & & & & & &55 &91$\pm$ 5 &1 \cr
& & & & & & &101 &90$\pm$11 &1 \cr
& & & & & & &146 &40$\pm$18 &1 \nl 

\tablebreak

NGC 4261 &0.17$\pm$0.015 
&159
&20 &55
&10/4 &11/4
&122 &22$\pm$ 9 &1 \cr
& & & & & & &0 &$-$76$\pm$20 &1,4 \cr
& & & & & & &62 &$-$23$\pm$15 &1 \cr
& & & & & & &90 &4$\pm$17 &1,4 \nl

NGC 4374 &14$\pm$0.02
&126
&10 &40
&\nodata &\nodata
&\nodata &\nodata &1\nl

NGC 4472 &0.17$\pm$0.005
&161
&30 &60
&10/4 &\nodata\tablenotemark{b}
&60 &$-$65$\pm$10 &1 \cr
& & & & & & &85 &$-$49$\pm$ 7 &5 \cr
& & & & & & &89 &$-$62$\pm$24 &1,4 \cr
& & & & & & &119 &$-$51$\pm$ 2 &1 \cr
& & & & & & &179 &4$\pm$ 6 &1,4 \nl

NGC 4486 &0.06$\pm$0.015
&161
&20 &60
&10/4 &\nodata\tablenotemark{b}
&9 &15$\pm$15 \cr
& & & & & & &69 &8$\pm$ 6 &1 \cr
& & & & & & &99 &$-$13$\pm$14 &1,4 \cr
& & & & & & &129 &$-$9$\pm$ 7 &1 \cr
& & & & & & &159 &$-$4$\pm$ 8 &1 \nl

NGC 4636 &0.12$\pm$0.03
&152
&15 &48
&\nodata &\nodata
&\nodata &\nodata &1\nl

NGC 4839 &0.29$\pm$0.07
&62
&4 &29
&\nodata &\nodata
&\nodata &\nodata &1\nl

NGC 7626 &0.11$\pm$0.01
&7
&5 &20
&9/4 &10/4
&28 &$-$25$\pm$26 &1 \cr
& & & & & & &87 &$-$8$\pm$14 &1 \cr
& & & & & & &118 &14$\pm$14 &1 \cr
& & & & & & &149 &0$\pm$ 9 &1 \cr
& & & & & & &178 &16$\pm$ 9 &1

\enddata

\tablenotetext{a}{Degrees from minor axis}
\tablenotetext{b}{Only one value of $k$ needed since the logarithmic
slope is almost constant between R$_{min}$ and R$_{max}$} 
\tablenotetext{c}{Sources for Kinematic Data:
(1) Davies \& Birkinshaw\ (1988)
(2) Franx {\it et al.}\ (1989)
(3) Binney {\it et al.}\ (1990)
(4) Bender {\it et al.}\ (1994)
(5) Fried \& Illingworth\ (1994)}
 
\end{deluxetable}

\begin{deluxetable}{rll}
\footnotesize
\tablecaption{Sample Anisotropies from Various Dynamical Assumptions
	\label{tbl-2}}
\tablewidth{0pt}
\tablehead{ \colhead{Case} & \colhead{Restriction} & $A_s$}
\startdata
NK & No kinematics, photometry only                & $0.10$ \cr
 0 & Maximal Ignorance (all models)                & $0.18$ \cr
 1 & $C(T,c)=$ ``Prescription 1''\tablenotemark{a} & $0.19$ \cr
 2 & $C(T,c)=$ ``Prescription 2''\tablenotemark{a} & $0.16$ \cr
 3 & $C(T,c)=$ ``Prescription 3''\tablenotemark{a} & $0.20$ \cr
 4 & $C(T,c)=$ ``Prescription 4''\tablenotemark{a} & $0.17$ \cr
 5 & $C=\infty$ (around short axis)                & $0.43$ \cr
 6 & $C=1$                                         & $0.17$ \cr
 7 & $C=0.5$                                       & $0.26$ \cr
 8 & $C=0$ (around long axis)                      & $1.17$ \cr
X1 & spheroidlike/spheroidlike\tablenotemark{b}    & $0.32$ \cr
X2 & disklike/disklike\tablenotemark{b}            & $0.20$ \cr
X3 & spheroidlike/disklike\tablenotemark{b}        & $0.21$ \cr
X4 & disklike/spheroidlike\tablenotemark{b}        & $0.13$ \cr
\enddata
\tablenotetext{a}{Prescriptions 1 -- 4 given by equations (11) -- (14)
	of Statler (1994b).}
\tablenotetext{b}{Form of $v^\ast(t)$ for long axis/short axis tubes.}
\end{deluxetable}

\newpage
\figcaption{
Stack-smooth-iterate algorithm in 1 dimension, finding the distribution of
the intrinsic quantity $X$ from measurements of an observable $x$. Likelihoods
$L_i(X)$ for 9 measured $x$ values ({\em a\/}) are multiplied by the initial
model parent distribution ({\em b\/}), summed ({\em c\/}), and smoothed to
produce an improved parent ({\em d\/}). Result after 20 iterations
({\em e, solid line\/}) is a decent match to the true parent distribution
({\em dotted line\/}). Estimates of $X$ values for individual objects
({\em f\/}) differ from the original likelihoods.
\label{f.oned}}

\figcaption{
($a$--$d$) Parent intrinsic shape distributions, in the space of
triaxiality $T$ of the mass distribution and flattening $c_L$ of the
light distribution, obtained under different
assumptions for the internal dynamics of the galaxies. In each panel,
round prolate galaxies are at upper left, flattened oblate galaxies
at lower right; objects in between are triaxial. Contours enclose
68\% and 95\% of the total probability. 
($a$) ``Maximal ignorance'' distribution, an unweighted combination of all
models; 
($b$) result from photometry only, omitting all kinematic data;
($c$) rotation solely around the short axis;
($d$) rotation solely around the long axis. 
($e$, $f$) Orientation distribution of the
sample, for the shape distributions in ($c$, $d$) respectively.
Centers of the circles correspond to views down the short axis, top and bottom
edges to views down the long axis, and extreme right-hand edge to views down 
the intermediate axis. These cases can be ruled out because of the
strong orientation bias implied for the sample.
\label{f.ignorant}}

\figcaption{
($a$) Triaxiality distribution in the maximal ignorance case, obtained by
integrating Figure \ref{f.ignorant}a
over flattening $c$. Dashed lines demarcate nearly
prolate (left), triaxial (center), and nearly oblate (right) regions.
Percentages indicate fractions of total probability in each region. 
($b$) Expected distribution of apparent ellipticities in the maximal
ignorance case ({\it smooth curve\/}). Histogram shows ellipticity
distribution for a sample of 165 galaxies from Ryden (1992). 
\label{f.ignorantoned}}

\figcaption{Posterior probability densities in the intrinsic shape plane
for each of the 9 galaxies which show significant rotation, 
using the maximal ignorance parent distribution.
Contours indicate the 68\% and 95\% highest posterior density regions.
\label{f.montage}}

\figcaption{Parent intrinsic shape distributions obtained under
different assumptions for the ``disklike'' or ``spheroidlike'' character
of the rotation in long-axis and short-axis tubes, as indicated.
\label{f.xmodels}}

\figcaption{Triaxiality distributions for the same four cases in
in fig. \ref{f.xmodels}, obtained by integrating the parent
distributions over $c$. Spheroidlike rotation in long-axis or
short-axis tubes suppresses the fraction of prolate or oblate objects,
respectively.
\label{f.xmodelsT}}

%%%%%%%%%%%%%%%%%%%%%%%%%%%%%%%%%%%%%%%%%%%%%%%%%%%%%%%%%%%%%%%%%%%%%%%%%%%%%%
%  This bit just prints all the figures at the end of the paper

\begin{figure}[p]{\hfil\epsfbox{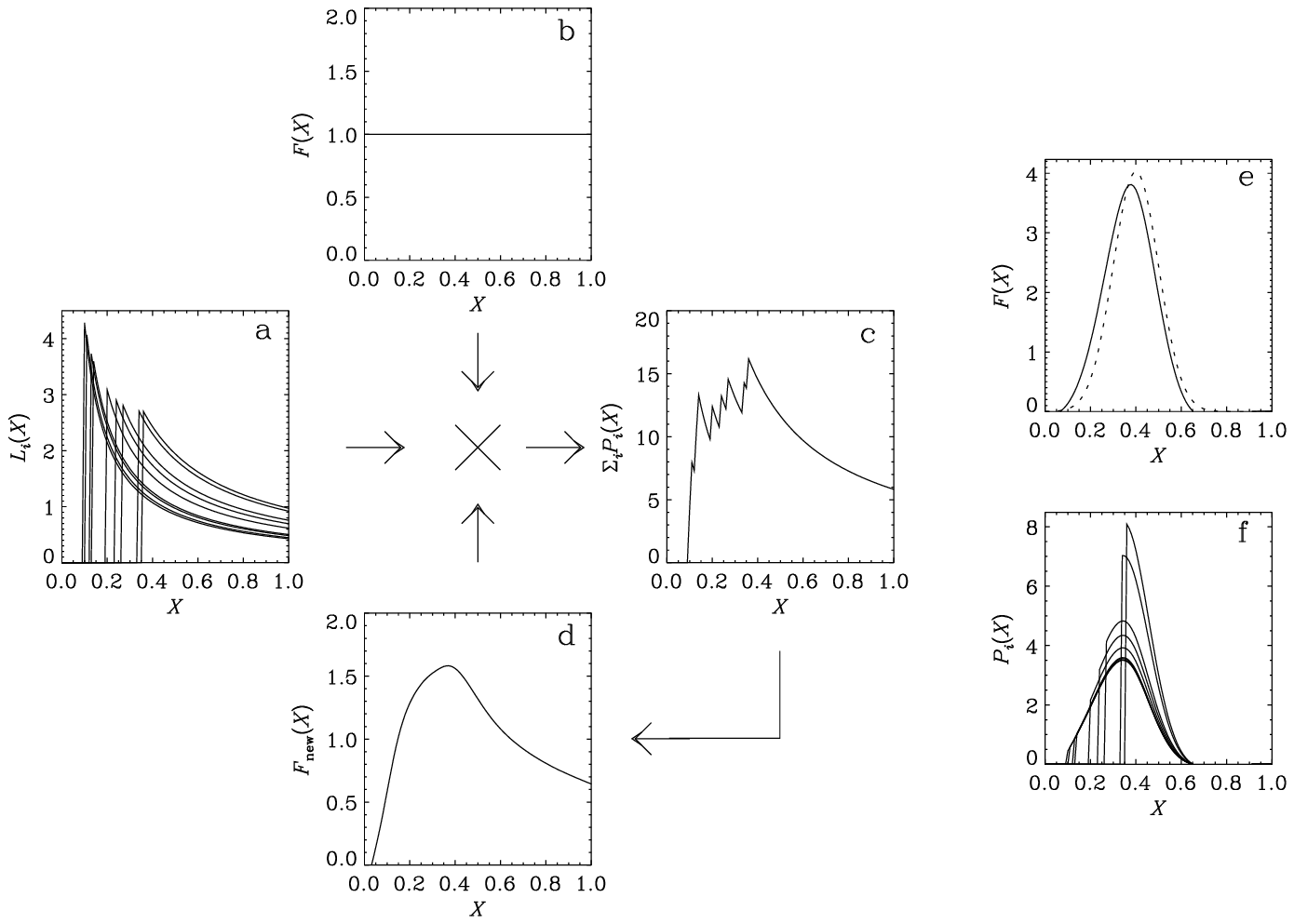}\hfil}\\
Fig. \ref{f.oned}
\end{figure}

\begin{figure}[p]{\hfil\epsfxsize=5.5in\epsfbox{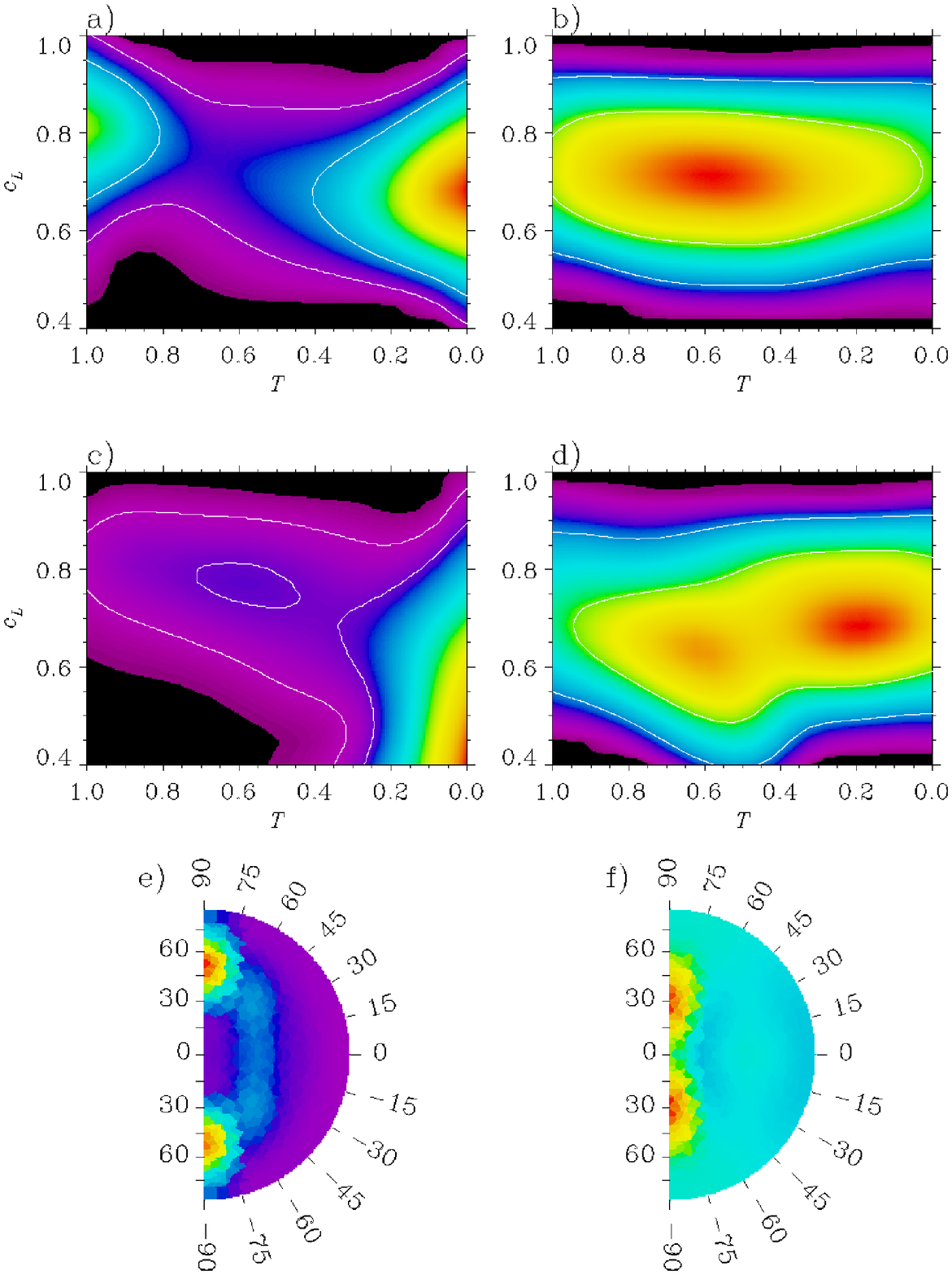}\hfil}\\
Fig. \ref{f.ignorant}
\end{figure}

\begin{figure}[p]{\hfil\epsfxsize=3.2in\epsfbox{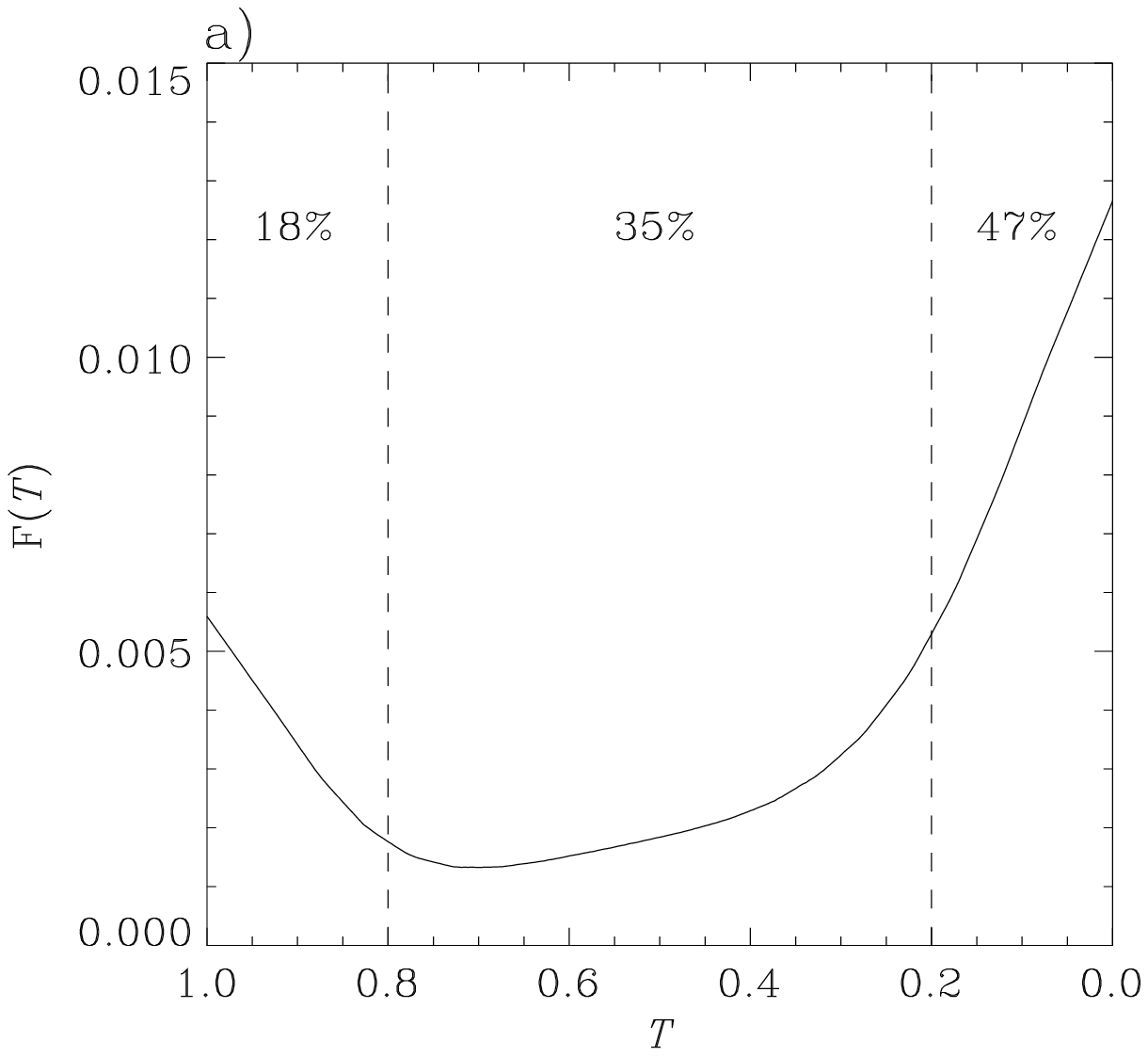}\hfil
\epsfxsize=3.2in\epsfbox{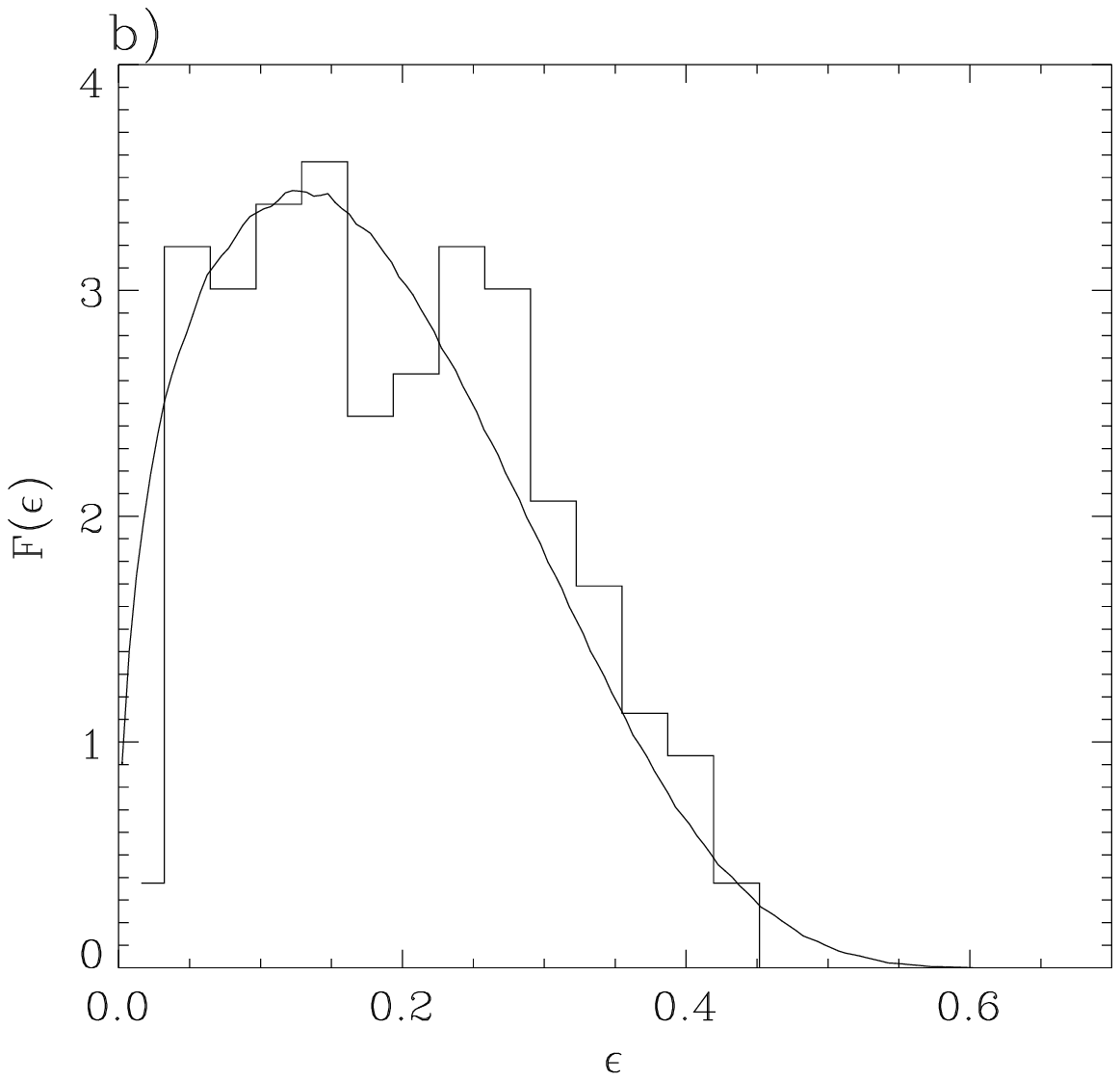}\hfil}\\
Fig. \ref{f.ignorantoned}
\end{figure}

\begin{figure}[p]{\hfil\epsfxsize=6.2in\epsfbox{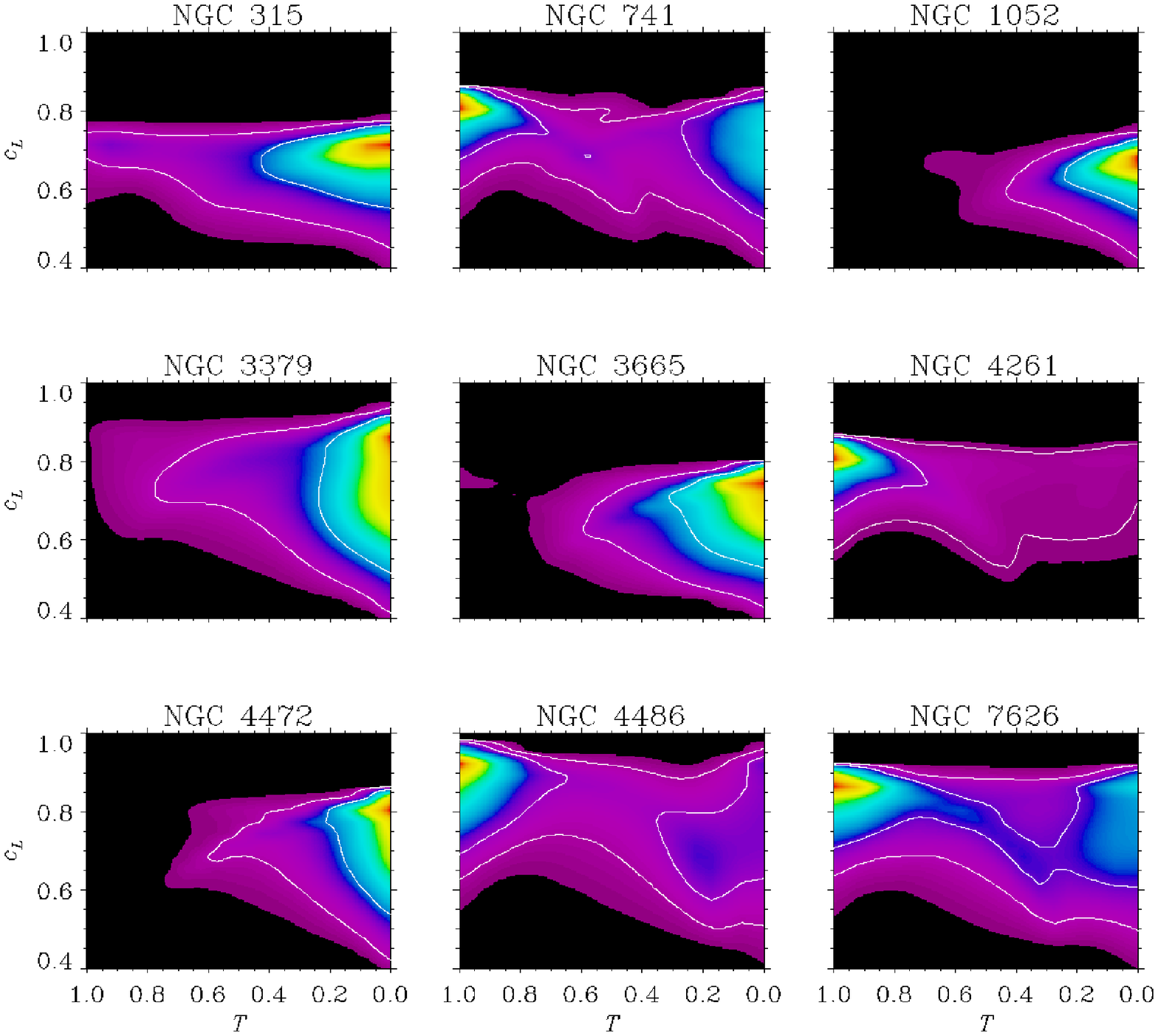}\hfil}\\
Fig. \ref{f.montage}
\end{figure}

\begin{figure}[p]{\hfil\epsfxsize=6.2in\epsfbox{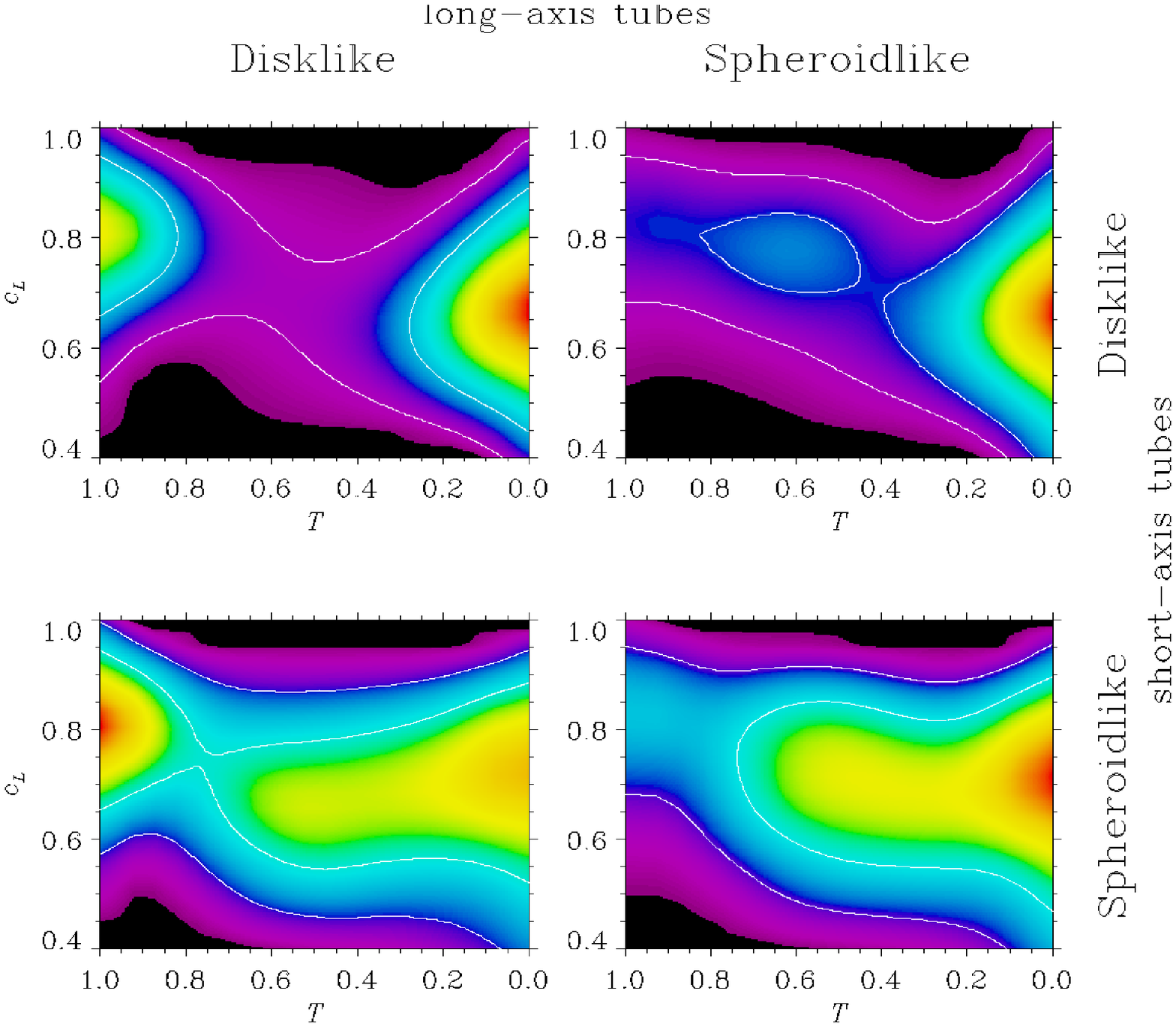}\hfil}\\
Fig. \ref{f.xmodels}
\end{figure}

\begin{figure}[p]{\hfil\epsfbox{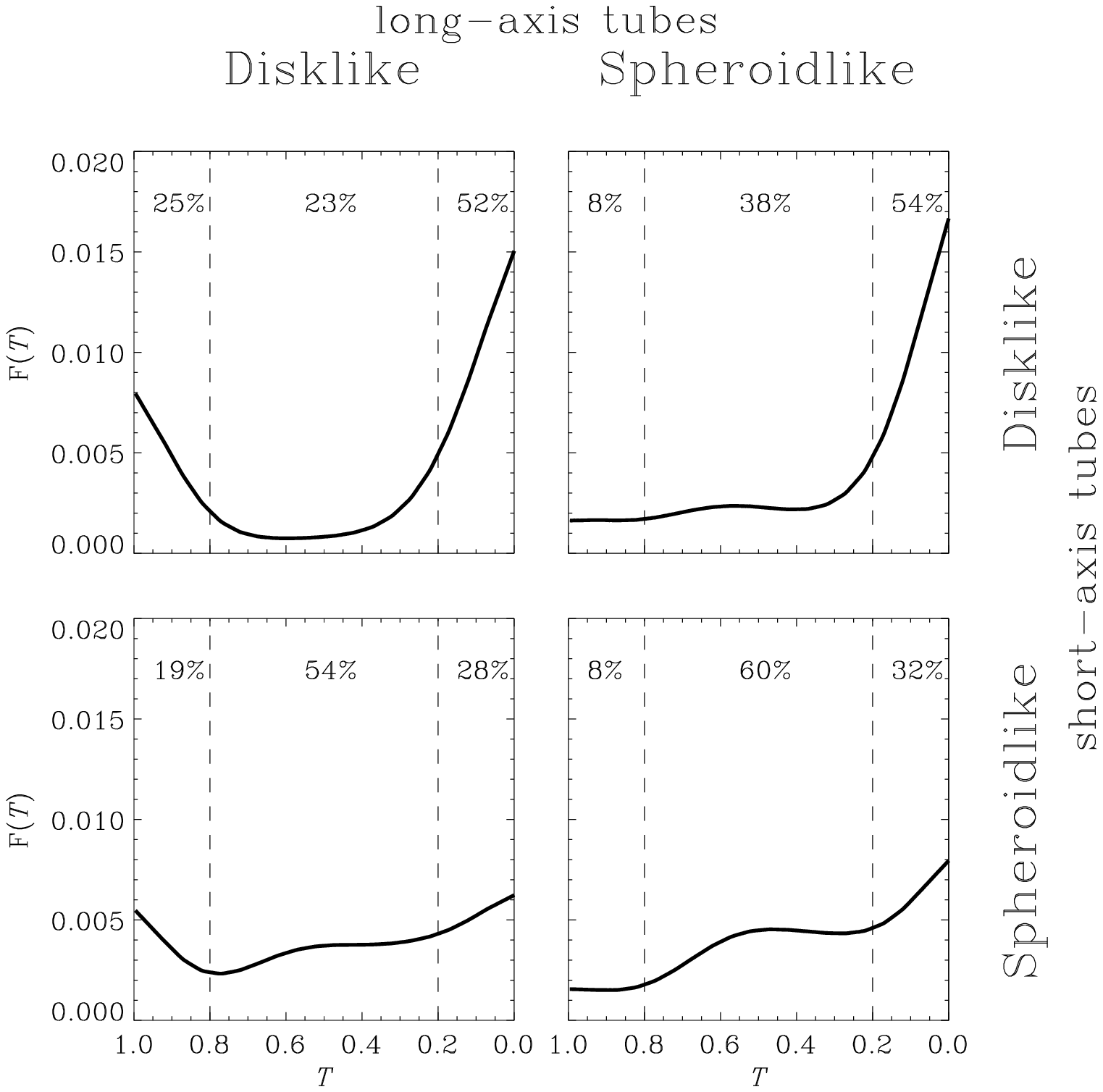}\hfil}\\
Fig. \ref{f.xmodelsT}
\end{figure}

%%%%%%%%%%%%%%%%%%%%%%%%%%%%%%%%%%%%%%%%%%%%%%%%%%%%%%%%%%%%%%%%%%%%%%%%%%%%%%

\end{document}